\documentstyle[twoside,12pt,epsf]{article}
\begin{document}
\centerline{\bf CMB EXPERIMENT BASELINE DETERMINATION}
\centerline{\bf BY MINIMIZING PIXEL OBSERVATIONS DISPERSION}
\bigskip
\noindent
George F. Smoot \\
LBNL, SSL, CfPA, Department of Physics \\
University of California \\
Berkeley CA 94720 \\
\bigskip

\centerline{\bf Abstract}
This paper discusses a couple of approaches to removing 
instrument gain and baseline variation from observations
of the CMB anisotropy.
These techniques were tested on the COBE DMR data, a balloon experiment, 
and by monte carlo simulations.
This work, though considered technical, is made available
because CMB anisotropy data processing is currently very active 
and a significant number of groups are working on this issue.
Specifically, time-ordered estimates of the gain and baseline
via fitting to limited sections of data or to the full 
data set are considered and compared.

A significant point is the there is a simple formulation
to estimate the pixel-to-pixel correlation added
by whatever baseline removal is utilized.

\section{Introduction}
Instruments used in making cosmic microwave background (CMB) 
sky maps (observations) have slow drifts in the signal levels
due to the change in instrument baseline and gain.
Data processing to remove this instrument signature and produce
calibrated data requires making an estimate of the baseline
and gain as a function of time and correcting the input data stream.
Typically, the baseline and gain are measured or estimated at intervals 
on the fly - i.e. in time sequence - and a smooth curve fitted through
these measurements for correction of the data.
This is calculationally efficient but carries some risk of removing
the wanted signal and perhaps worse introducing angular correlations
in the resultant data product, since sky scanning couples time
dependence to angles on the sky. 

An alternate approach is to collect the full data set and perform
a simultaneous and complete fit to the resultant map and baseline/gain
by minimizing the dispersion of the repeated observations
of each pixel and the independent estimates of baseline and gain.
Ideally this is the most powerful technique and least likely
to introduce large scale angular correlations.
A small amount of angular correlation can be produced simply
due to the random errors of observations being over-subtracted
in the pixel variance minimization. 
In practice this technique works well with observation schemes
that provide many repeated observations of a significant number
of pixels on many time scales.
This technique is computationally intensive.

Some tests that have been used are discussed.
This paper is primarily based upon COBE Note 5047 which I wrote
in 1991 and revised in November 1992.
The topic has once again become of significant interest
so that I have added some material and included the bulk information
of the COBE Note. A few changes must be made by me or by the reader
since some observation schemes have a single beam and thus minimizing
each single pixel observations' dispersion is relevant.
Other experimental schemes, such as the COBE DMR and MAP, 
observe the difference between two pixels.
In the case of the COBE DMR we then minimize the dispersion
in the observations of each pixel-pair (or sometimes called pixel-perms).
The basic concept remains the same but the bookkeeping in the 
data processing is somewhat different.

\section{Concept}
The basic idea is that the instrument baseline and gain are
slowly changing effects that can be described by a function
of time with few parameters compared to the number of data samples taken.
Ensuring this is a critical aspect of a successful experiment design.

In the case of the COBE DMR,
the standard approach is to fit the baseline 
to the averages of the data over a long time interval sample.
For example, average the data for 60 to 100 complete rotations of
the spacecraft.
For its accuracy this procedure depends upon the rotation of the spacecraft 
and the essentially small true signal so that data average is 
the actual baseline.

The major assumption of this approach is that the rotation interchanges 
the antennas reversing the true sky signal leaving the baseline unchanged.
This is only an approximation. Why not treat it this way explicitly?
That is why not find all the symmetric measurements and fit to the baseline
that leaves the true sky signal antisymmetric under horn antenna interchange?
This can be done by taking all the observations of each pixel pair and
choosing a baseline that minimizes the dispersion in pixel pair observations 
values around the average for each pixel pair.

Even using pixel (pixel pairs) is a small approximation which could be
noticeable unless the pixel size is much smaller than the beam. 
That is, this approach assumes that the signal observed for the pixel
does not have a significant change from one side of the pixel
to the other. 
We minimize the effect of signal gradient on the COBE DMR data processing 
by both removing the CMB dipole signal
and by using higher angular resolution pixels for the Galactic plane region
where there was a significant varying signal.
In prinicple one could do this by subtracting a model
of the sky from the input data and iterating to get a `best-estimate model'
and a map of the instrument noise.

\section{Mathematical Treatment}

The time ordered DMR data - temperature differences -  
have a slow drift which must be removed before a map is produced. 
In the standard data processing this drift is largely removed 
by a high-pass filter. 
We call this the baseline subtraction and it is accomplished using
spline (and/or polynominal) fits to the time-ordered data.
For the DMR both methods were used and compared.
This process actually introduces some correlation from measurement
to measurement, adds noise, and removes some of the true signal
since the baseline fitting includes the noise and the signal
under the not quite true assumption that they average to zero over the data
sampled for the baseline determination.

Using the alternate technique of global fitting
the temperature differences can be recovered and any residual
drifts removed by fitting the data to a spline or similar function to minimize 
the {\it variance} in each sky pixel. 
The MIT group used this procedure for their single-horn experiment
maps where drifts were particularly troublesome
(Cottingham 1987, Boughn {\it et al.} 1992). 

The mathematical treatment is relatively straightforward.
The approach can be simplified if we can find a formulation
that describes the original or the residual baseline drift as some
linearized time dependent series:
$$ Drift(t_i) = \sum_{k=1}^{N_{k}}a_{k}B_{k}(t_{i}) $$
where $B_{k}(t_{i})$ is a function in the time series - e.g. a polynominal
in $t_i$ or a spline - and $a_k$ is the coefficient for that
time series function. 
Using this formulation approach makes the
dependence of the equation linear.
After this we will call it a spline fit but any set of functions
$B_{k}(t_{i})$ that scale by amplitude will allow matrix manipulation
to find the minimum $\chi^{2}$.

The $\chi^{2}$ for the fit is given by:

$$ 
\chi^{2} 
= \sum_{all~t_i}
\biggl( S(t_{i}) - Drift(t_i) - \overline{S_{i}} \biggr)^2 /\sigma_i^2
= \sum_{j=1}^{N_{\rm pix}}\sum_{i=1}^{N_{j}}
\biggl( S_{j}(t_{i}) - 
\Bigl( \sum_{k=1}^{N_{k}}a_{k}B_{k}(t_{i}) 
+ \overline{S_{j}} \Bigr)\biggr)^{2} /\sigma_i^2,
$$
where $S_j$ is the measured temperature difference 
which is subdivided into $S_j(t_i)$, 
the temperature difference of pixel pair $j$, 
and the mean of the regressed signal, ${\rm\overline{S_{j}}}$, 
in a pixel-pair is given by: 
$${
\overline{S_{j}}={1\over N_{j}}\sum_{l=1}^{N_{j}}
\biggl( S_{j}(t_{l}) -  \sum_{k=1}^{N_{k}}a_{k}B_{k}(t_{l}) \biggr).
}$$
 In these expressions, 
${\rm N_{pix}}$ is the total number of pixel-pairs used in the fit;
${\rm N_{j}}$ is the number of data points for pixel-pair j ;
${\rm S_{j}(t_{i})}$ is the high-passed signal at time ${\rm
t_{i}}$ in pixel pair j; 
for a spline fit 
${\rm N_{k}}$ is the number of fit parameters, one for
           each of the B--spline coefficients; 
${\rm a_{1}~to~a_{k}}$ are the fit parameters; and
${\rm B_{k}(t_{i})}$ is the ${\rm k^{th}}$ B--spline evaluated
           at time ${\rm t_{i}}$.

Using our standard procedure we can vary $a_k$ and ${\rm\overline{S_{j}}}$
to get the minimum chi-squared ($\chi^2$) or 
in the linear formulation we can get a set of normal equations
by taking the derivative of the chi-squared with respect to those parameters.
If the data weights are the same for each measurement we can set
$\sigma_i = 1$ and renormalize later.
The linear equations are:
$$\pmatrix{
N_{1}&0&\ldots&\sum\limits_{i=1}^{N_1}B_{1}(t_{i})&
\sum\limits_{i=1}^{N_1}B_{2}(t_{i})&\ldots\cr 
0&N_{2}&\ldots&\sum\limits_{i=1}^{N_2}B_{1}(t_{i})&
\sum\limits_{i=1}^{N_2}B_{2}(t_{i})&\ldots\cr 
\vdots&\vdots&\ddots&\vdots&\vdots&\vdots\cr 
\sum\limits_{i=1}^{N_1}B_{1}(t_{i})&\sum\limits_{i=1}^{N_2}B_{1}(t_{i})&\ldots&
\sum\limits_{all~t_i}B_{1}^{2}(t_{i})&
\sum\limits_{all~t_i}B_{1}B_{2}&\ldots\cr 
\sum\limits_{i=1}^{N_1}B_{2}(t_{i})&\sum\limits_{i=1}^{N_2}B_{2}(t_{i})&\ldots&
\sum\limits_{all~t_i}B_{1}B_{2}&
\sum\limits_{all~t_i}B_{2}^{2}(t_{i})&\ldots\cr 
\vdots&\vdots&\vdots&\vdots&\vdots&\ddots\cr }
\pmatrix{
{\rm\overline{S_{1}}} \cr
{\rm\overline{S_{2}}} \cr
\vdots \cr
a_1 \cr
a_2 \cr
\vdots \cr}
=
\pmatrix{
\sum\limits_{i=1}^{N_1}S_{1}(t_{i}) \cr 
\sum\limits_{i=1}^{N_2}S_{2}(t_{i}) \cr 
\vdots \cr
\sum\limits_{all~t_i}B_{1}(t_{i})S(t_i) \cr
\sum\limits_{all~t_i}B_{2}(t_{i})S(t_i) \cr
\vdots \cr
}$$
where, as above,  the ${\rm N_{j}}$ are the number of data points
for pixel-pair j. This matrix equation can be written as:
$$\pmatrix{
\alpha&\beta\cr
\beta^{\dag} &\gamma\cr }
\pmatrix{
{\rm\overline{S_j}} \cr
a_k \cr }
=
\pmatrix{
\sum_{pixel~pair_j}S_j(t_i) \cr
\sum_{all~t_i} B_n (t_i) S(t_i) \cr
}$$
where $\alpha$ is the $N_{pix}$ by $N_{pix}$ diagonal matrix
containing the number of times each pixel pair was sampled. The
matrix $\beta$ has dimension $N_{pix}$ by $N_{k}$ and $\gamma$
has dimension $N_{k}$ by $N_{k}$ where $N_{k}$ is the number of
spline coefficients. 
The upper part ($\alpha$), relating to
$N_j {\rm\overline{S_j}} = \sum_{pixel~pair_j}S_j(t_i)$, 
is just our usual averaging together of the
pixel-pair observations to find the mean.
The lower part ($\gamma$) is just the linear set of equations used to
find the spline fit to the baseline.
The corners ($\beta$ and $\beta^{\dag}$) are the coupling of the
signal and the baseline which we normally ignore 
in the standard data processing by trying to have
the average pixel-pair difference small - both by removing a model
of the signal and by having equal number of observations
with the horns in one orientation and the exchanged orientation.

The curvature matrix can be inverted to find the
covariance matrix of the fit. 
The pixel-pair covariance part of the matrix is given by
$${C=\alpha^{-1} + \alpha^{-1}\beta
(\gamma-\beta^{\dag}\alpha^{-1}\beta)^{-1}\beta^{\dag}\alpha^{-1}.
} \eqno(1)$$
This is an $N_{pix}$ by $N_{pix}$ matrix. 
This is relatively easy to calculate because the upper left section, $\alpha$,
is diagonal and the inverse is just the inverse of the diagonal elements.
Thus though in principle it is a huge matrix to invert
there is little inversion necessary.

The solution to the matrix equation would be the pixel-pair temperature 
differences and the spline coefficients.
The solution for the pixel-pair temperature differences
is just the average of the high pass filtered data
with a small correction for the best-fit baseline - namely
the second portion = off-diagonal elements of the covariance matrix.
We would have to have the full matrix to find the off-axis portion;
however, we can make a rough estimate of these values.
For section the upper left section
$N_{pix} = \approx 60 x 6144$ and $N_{k} = ?$. 
We can estimate a typical value of the number of observations on a single pixel pair
is roughly 100.
Now we need an estimate of the number of terms in the spline.
Let us assume that the knots are set effectively at one per hour
for a total of 364 x 24 = 9000 per year
or about one per 7200 data points.
The predicted mean value of the off-diagonal terms is
$2.5\times 10^{-5}{\rm mK^{2}}$ (0.005 mK).

The covariance matrix is then `normalized' to produce the correlation matrix.

\section{Fit-induced pixel-to-pixel correlations.}

How much correlation does an imperfect baseline subtraction introduce 
into the data, especially on large angular scales? 
The correlations could be checked for by injecting a `hot' pixel into
the time ordered data stream and finding where it ends up in the map. 
This was done for a few pixels and only small correlations were found. 
A better method is to find the correlation matrices for 
various fits. 

Dave Cottingham devised this way to compute the correlations
between pixels introduced by this fit for his balloon-borne instrument map. 
Formally, the number of fit parameters is the number of spline coefficients. 
However, in the expression above one may also view 
the mean temperature difference between each pixel-pixel as a fit parameter. 
If one then gets the normal equations it is easy using the formula (1)
to compute the pixel to pixel correlation matrix from the correlation
matrix shown above.

\section{Gain Drift}

If the gain is slowly varying with time, then the straight baseline
subtraction will alias the data slightly.
We could also use the minimizing of pixel-pair variance for this too.
This is not the most effective way to remove the gain drift.
A better way is to use the pixel-pair temperature differences for the
dipole and the moon signal - particularly the dipole.
One can then either treate this separately or in an even larger 
combined fitting.

\section{Tests of calibration.}  

One way to check the effects of section offset and gain
errors is to produce maps that have the offsets and gains 
vary slowly with time.
To introduce ersatz large scale structure into the 
correlation function of the magnitude observed requires altering
the relative baselines by about 3 to 4 sigma (eg. by
adding ${\rm 100\mu K}$). 
It is harder to change the shape by adjusting the gain.

\section{Applying this to the Pixel Values.}  

I speculated that this treatment of the baseline could be built 
into the program which produces the sky map. 
It would make the normal equations much larger.
We have the same chi-squared equation to minimize but we can introduce
the pixel temperature values instead of the pixel-pair values.
For the COBE DMR the sky map generating equations
are more complicated than the pixel-pair equations,
since there are non-diagonal terms connecting pixels
since it is a differential experiment.

The $\chi^{2}$ for the fit is given by:

$$ 
\chi^{2} 
= \sum_{all~t_i}
\biggl( S(t_{i}) - Drift(t_i) - T_j + T_k \biggr)^2 /\sigma_i^2
$$
where $S(t_i)$ is the measured temperature difference at time $t_i$,
$T_j$ is the temperature of pixel j.
In these expressions, 
${\rm N_{pix}}$ is the total number of pixel used in the fit;
${\rm N_{j}}$ is the number of data points for pixel j ;
${\rm S(t_{i})}$ is the high-passed signal at time ${\rm
t_{i}}$; 
for a spline fit 
${\rm N_{k}}$ is the number of fit parameters, one for
           each of the B--spline coefficients; 
${\rm a_{1}~to~a_{k}}$ are the fit parameters; and
${\rm B_{k}(t_{i})}$ is the ${\rm k^{th}}$ B--spline evaluated
           at time ${\rm t_{i}}$.

Using our standard procedure we can vary $a_k$ and $T_j$ and $T_k$
to get the minimum chi-squared ($\chi^2$) or get a set of normal equations
by taking the derivative of the chi-squared with respect to those parameters.
If the data weights are the same for each measurement we can set
$\sigma_i = 1$ and renormalize later.
The linear equations are:
$$\pmatrix{
\alpha&\beta\cr
\beta^{\dag} &\gamma\cr }
\pmatrix{
{T_j} \cr
a_k \cr }
=
\pmatrix{
\sum_{pixel_j}S(t_i) \cr
\sum_{all~t_i} B_n (t_i) S(t_i) \cr
}$$
where $\alpha$ is the $N_{pix}$ by $N_{pix}$ sparse near-diagonal matrix
we use in the normal equations for making maps from differential measurements.
Matrix $\beta$ has dimension $N_{pix}$ by $N_{k}$ and $\gamma$
has dimension $N_{k}$ by $N_{k}$ where $N_{k}$ is the number of
spline coefficients. 
The upper part ($\alpha$), relating to
is just our usual normal equations for taking the
pixel-pair observations to find best fitted map.
The lower part ($\gamma$) is just the linear set of equations used to
find the spline fit to the baseline.
The corners ($\beta$ and $\beta^{\dag}$) are the coupling of the
signal and the baseline which we normally ignore by trying to have
the average pixel-pair difference small - both by removing a model
of the signal and by having equal plus and minus observations.

The curvature matrix can be inverted to find the
covariance matrix of the fit. 
The pixel covariance part of the matrix is given by
$${C=\alpha^{-1} + \alpha^{-1}\beta
(\gamma-\beta^{\dag}\alpha^{-1}\beta)^{-1}\beta^{\dag}\alpha^{-1}.
}$$
This is an $N_{pix}$ by $N_{pix}$ matrix. 

Conclusion: This shows that we can correct for the baseline effects
and estimate the pixel-to-pixel correlation either in forming the pixel
perms before they go into map fitting software 
or in making the map so long as we keep
the time information or the sum vector $B_k(t_i)S(t_i)$ for each pixel
or pixel pair.

This approach is computationally much more intensive and requires
being able to solve large matrix equations. For COBE DMR we determined
that it was not worth the additional complexity since the instrument
was so very well behaved.

\section{Additional Work}
Since this original work, a fair interest has developed in the question
of striping in CMB maps.
It is driven in large part by the studies for proposals and the development
of future satellite missions. 
It is relevant for balloon-borne experiments now underway.
The early focus was upon finding a method to estimate the
map striping that would be produced by $1/f$ noise 
(a good estimate of the expected baseline drift on intermediate time scales)
as a function of scan strategy and speed
for both bolometer and HEMT-amplifier receivers (Janssen et al. 1996).

The scan strategy, particulary the time scales upon which one revisit pixels,
is critical in equal part with the stability, i.e. power spectrum of 
instrument noise, of the observing instrument.
It should be borne in mind that the map reconstruction software
may also reduce or produce striping. 
Some approaches produce correlations between pixels even if the
observations did not.
A comparison of scan strategies with minimally intesecting great circles
and fairly maximally cross-linked smaller circles (Wright 1966)
indicates cross-linking is desirable, does produce some striping
and has generated futher controversey.
Some of the controversey has to do with different configurations
and assumed noise and with different approaches (e.g. J. Delabrouille 1997).

An important conclusion is that the specific experiment/observation
must be carefully simulated and tests applied to understand the
implications of the instrument noise, scan strategy, and make
construction algorithm. 
There are rules of thumb and analytic estimators that provide guidance
but the full treatment is necessary for the final evaluation.
It appears that several approaches can be made to work with sufficient
accuracy provided the instrument and observing enviroment
are adequate.

\bigskip
\centerline{References.}  
\bigskip

Boughn, S., {\it et al.} 1992, AP. J. {\bf 349}, L49.

Cottingham, D., {\it A Sky Temperature Survey at 19.2 GHz
Using a Balloon Borne Dicke Radiometer for Anisotropy Tests of the
Cosmic Microwave Background} Ph.D. Thesis, Princeton University, 
1987. 

Delabrouille, J. "Analysis of the accuracy of a destriping method
for future CMB mapping with the PLANCK SURVEYOR satellite."
submitted A \& A 1997.

Janssen, M. et al. astro-ph 9602009 "Direct Imaging of the CMB from Space"
submitted to Ap. J. 

Wright, E. astro-ph/9612006 "Scanning and Mapping Strategies for CMB 
Experiments"

\vfil\eject
\end{document}